\documentclass{article}

\usepackage{amsmath, amssymb, amsthm, authblk, bbm, booktabs, graphicx, setspace, subcaption, tikz}

\usepackage{enumitem}
\usepackage{apacite}
\usetikzlibrary{positioning}

\usepackage[margin = 1in]{geometry}

\title{A regression framework for a probabilistic measure of cost-effectiveness}
\author{Nicholas Illenberger\textsuperscript{1}, Nandita Mitra\textsuperscript{1}, Andrew J. Spieker\textsuperscript{2}}
\date{\small{\textsuperscript{1}\textit{University of Pennsylvania, Department of Biostatistics, Epidemiology, and Informatics, Philadelphia PA}} \\
\small{\textsuperscript{2}\textit{Vanderbilt University Medical Center, Department of Biostatistics, Nashville TN}}}

\begin{document}

\maketitle

\begin{abstract}
To make informed health policy decisions regarding a treatment, we must consider both its cost and its clinical effectiveness. In past work, we introduced the net benefit separation (NBS) as a novel measure of cost-effectiveness. The NBS is a probabilistic measure that characterizes the extent to which a treated patient will be more likely to experience benefit as compared to an untreated patient. Due to variation in treatment response across patients, uncovering factors that influence cost-effectiveness can assist policy makers in population-level decisions regarding resource allocation. In this paper, we introduce a regression framework for NBS in order to estimate covariate-specific NBS and find determinants of variation in NBS. Our approach is able to accommodate informative cost censoring through inverse probability weighting techniques, and addresses confounding through a semiparametric standardization procedure. Through simulations, we show that NBS regression performs well in a variety of common scenarios. We apply our proposed regression procedure to a realistic simulated data set as an illustration of how our approach could be used to investigate the association between cancer stage, comorbidities and cost-effectiveness when comparing adjuvant radiation therapy and chemotherapy in post-hysterectomy endometrial cancer patients.
\end{abstract}

\section{Introduction}
\par Cost-effectiveness analyses are a useful tool for aggregating information on the differences in cost and clinical effectiveness between two comparator health interventions. Typically, a cost-effectiveness analysis relies on defining a particular kind of summary measure, which may be used to optimize the tradeoff between treatment costs and effectiveness. Historically, most analyses have involved either the incremental cost-effectiveness ratio (ICER) or the equivalent net monetary benefit (NMB). Each of these compares the mean difference in cost and the mean difference in some clinical measure between treatments. Specifically, the ICER is defined as the ratio of these two quantities and therefore has units of cost per unit of the clinical measure (for instance, \$USD per year, if the clinical outcome is marked by post-treatment survival time). The experimental treatment is declared cost-effective if the ICER lies below a certain threshold, known as the willingness-to-pay (WTP). The NMB was introduced by Stinnet et al. \cite{stinnett1998net} as a refined, yet equivalent measure of cost-effectiveness that suffers from neither a singularity in the denominator nor the known poor statistical properties associated with ratio quantities.
\par When analyzing medical costs that are accumulated over time, censoring poses specific barriers to straightforward estimation of mean costs. Correlation between costs at time of censoring and those at end-of-study precludes the use of methods which rely on the non-informative censoring assumption. This challenge is most commonly addressed by inclusion of inverse probability of censoring weights (IPCW) \cite{bang2000estimating}. Li et al. expanded this approach with a doubly robust method to account for exposure-outcome confounding in settings where data are derived from observational studies \cite{li2018doubly}.
\par In practice, the extent of cost-effectiveness associated with a treatment may vary across levels of an observed covariate. As an example, consider the setting of endometrial cancer, in which treatment often involves total hysterectomy. Following surgery, there is a large degree of variation in the treatment a patient receives, with some receiving adjuvant radiation, some receiving adjuvant chemotherapy, and some receiving neither \cite{latif2014adjuvant}. While costs associated with radiation and chemotherapy are certain to exceed those associated with a control condition, the survival benefit associated with either radiation or chemotherapy relative to neither could reasonably be expected to vary by patient covariates, e.g., cancer stage. Specifically, those with a lower cancer stage (and therefore, a lower overall risk of recurrence) may not experience the same degree of survival benefit from adjuvant treatment as compared to those with a higher cancer stage.
\par In previous work, regression methods for NMB have been developed to compare cost-effectiveness across subgroups defined by observed covariates. Willan et al. \cite{willan2004regression} show how NMB may be characterized as a function of the parameters in a regression of cost and effectiveness outcomes on treatment and observed covariates. Under the assumption of bivariate normal errors, the authors proposed a hypothesis test for the effects of covariates on NMB. Because cost distributions are typically skewed, Nixon et al. presented an alternative approach which could accommodate more flexible error structures \cite{nixon2005methods}. While positing regression models similar to those of Willan et al. \cite{willan2004regression}, their method implements Markov Chain Monte Carlo methods to accommodate gamma and log-normal distributed errors. These approaches are useful if we are interested in how the cost-effectiveness of an intervention (as measured by NMB or ICER) varies across levels of measured covariates.
\par In recent work, Spieker et al. \cite{spieker2019net} introduced a novel probabilistic measure of cost-effectiveness, the net benefit separation (NBS). The NBS characterises the stochastic ordering of individual net benefits (INB) between treated and untreated populations. For a given WTP, INB is defined as the difference between the cost a payer is willing to incur for an observed effectiveness measure and that which is truly incurred. Compared with NMB and ICER, which are functions of the average difference in cost and effectiveness measures between treatments, NBS measures the probability that a patient receiving treatment will experience greater treatment benefit than a patient receiving control. Because the distribution of medical costs, and consequently INBs, are often skewed, cost-effectiveness comparators that do not rely on the average values of cost or measures of effectiveness can be useful. In settings where treatment is randomized, NBS can be estimated nonparametrically using a scaled variant of the Wilcoxon rank-sum statistic.
\par As with NMB and ICER, extensions of NBS to settings with informative cost censoring and confounding have been proposed. Spieker et al. \cite{spieker2019net} introduce a semiparametric Monte Carlo standardization procedure to estimate NBS while adjusting for differences in the distribution of confounding variables across treatment arms. Models used within the standardization procedure may be fit using IPCW techniques if there are censored cost data. However, no methods currently exist which can explain variability in NBS using covariates. In this paper, we introduce regression methodology to account for variability in the NBS using observed covariates. In previous work, Pepe et al. developed regression methods for measures of stochastic ordering for use in diagnostic testing settings \cite{pepe1997regression}. Alonzo and Pepe later extended this work to improve computational efficiency in large samples \cite{alonzo2002distribution}. Because the NBS is defined as the stochastic ordering of INBs between treatment groups, we are able to embed their approach within our proposed methodology. We also describe how Monte Carlo standardization and IPCW methods may be used to account for censoring and confounding when performing NBS regression. Additionally, we propose a hypothesis testing framework for determining the strength of evidence for the effects of covariates on NBS.  
\par The remainder of this paper is organized as follows: In Section \ref{sec:nbs-review} we review the net benefit separation as proposed by Spieker et al. \cite{spieker2019net}.  In Section \ref{sec:regression} we define a conditional variant of the NBS and introduce our regression-based estimation procedure. In Section \ref{sec:simulations} we conduct simulation studies to examine the finite sample properties of this estimator. In Section \ref{sec:data}, we use our proposed approach to analyze the cost-effectiveness of adjuvant treatments for endometrial cancer. In the final section, we summarise the utility of our proposed cost-effectiveness parameters and discuss limitations of our methodology. 

\section{Net benefit separation}\label{sec:nbs-review}

\par Assume we have individual patient data on costs and treatment effectiveness. Let $A$ be a binary treatment indicator, $Z$ be a measure of effectiveness, $Y$ be a measure of medical cost, and let $i = 1,\hdots, n$ index sample units. In a typical setting, $Z$ and $Y$ may represent survival time and medical cost in USD(\$). For predefined WTP, $\lambda$, INB is defined as $B(\lambda) = \lambda Z - Y$. We assume without loss of generality that larger values of $Z$ are desirable, so that greater INB indicates greater treatment benefit. If we let $B_a(\lambda)$ represent a randomly sampled INB from a hypothetical population receiving treatment $A=a$, then NBS is defined as

\begin{equation*}
    \theta(\lambda) = P\big(B_1(\lambda) > B_0(\lambda)\big).
\end{equation*}

NBS can be interpreted as the probability that a randomly sampled unit from a population receiving treatment will have a greater INB than an independently sampled unit from a control population. For a particular $\lambda$, if $\theta(\lambda) = 0.5$, then the treatments are equally cost-effective with respect to NBS. Note that $\theta(\lambda) > 0.5$ does not imply that $\mathbb{E}\left[B_1(\lambda)\right] > \mathbb{E}\left[B_0(\lambda)\right]$, i.e. cost-effectiveness with respect to NBS does not imply greater expected INB. For example, if the distribution of INBs is skewed, then the expected benefit under control may be larger than that under treatment even if the majority of patients receiving treatment experience greater benefit than those receiving control. Thus, NBS can help provide a more complete view of the range of patient experiences, particularly when cost and effectiveness metrics are skewed.
\par For simplicity, we first consider estimation in a setting where treatment is randomized and with no censoring. Here, the observed distribution of INBs in the treatment and control arms form nonparametric estimates of the distributions of $B_1(\lambda)$ and $B_0(\lambda)$. Thus, NBS can be consistently estimated using a scaled variant of the Wilcoxon rank-sum statistic

\begin{equation*}
    \widehat{\theta}(\lambda) = \frac{1}{2 N_0}\left( \frac{1}{N_1} \sum_{i=1}^N 2 A_i R_i(\lambda) - N_1 - 1 \right),
\end{equation*}

where $N_a$ is the number of participants receiving treatment $A=a$ and $R_i(\lambda)$ is the pooled rank of the INB for the $i^{\text{th}}$ individual. Spieker et al. proposed a flexible standardization approach to estimate the NBS in the presence of measured confounders \cite{spieker2019net}. By modelling the joint distribution of observed variables, they are able to apply Monte Carlo methodology to sample INBs from treated and untreated populations which have the same underlying distribution of confounders. Applying the scaled Wilcoxon rank-sum statistic to the generated INBs provides an estimate of NBS that accounts for measured confounding. If data are censored, then IPCW can be incorporated in the modeling step to account for censored cost data. In the section that follows, we will extend this methodology to evaluate the association between measured covariates and the NBS.

\section{Regression of net benefit separation}\label{sec:regression}

 In this section we introduce regression methodology for NBS. Section \ref{sec:covarnbs} defines a conditional variant of NBS and describes procedures for estimation and hypothesis testing when treatment is randomized. In Section \ref{sec:standardization} we propose a Monte Carlo standardization procedure to address measured confounding. Section \ref{sec:ipcw} discusses how to incorporate IPCW to account for informative cost censoring.

\subsection{Conditional net benefit separation}\label{sec:covarnbs}
\par Suppose $X$ is some covariate of interest which may explain variation in the NBS. In an endometrial cancer setting, $X$ may represent cancer stage or age at diagnosis. We want to assess how the cost-effectiveness of a treatment changes with respect to $X$. Specifically, we may seek to (1) characterize the association between $X$ and NBS, (2) assess the NBS across levels of $X$, and (3) evaluate the strength of evidence for association. When treatment is randomized and $X$ is discrete, the first two goals can be accomplished by estimating $\theta(\lambda|X=x)$ using the scaled Wilcoxon rank-sum statistic within subsets defined by $X=x$. On the other hand, if $X$ is continuous, then this is not possible without first discretizing $X$. In either case, there are currently no methods to assess the strength of association between $X$ and the NBS. In this section, we introduce a procedure that can accomplish all three of the stated goals for continuous or discrete $X$.

For a given $\lambda$, the NBS conditional on $X$ is defined as
\begin{equation}
    \theta(\lambda|X) = P(B_1(\lambda) > B_0(\lambda) | X).
\end{equation}
It can be shown that the conditional NBS is equivalent to
\begin{equation}
    \theta(\lambda|X) = \int_0^1 S_{1|X}(S_{0|X}^{-1}(\omega)) d\omega
\end{equation}
where $S_{a|X}(u) = P(B_a(\lambda) > u|X)$ denotes the conditional survivor function for INB in a hypothetical population receiving treatment $A = a$ and $S_{a|X}^{-1}(u)$ is the quantile function for the same population. To estimate $\theta(\lambda|X)$, we build methodology parallel to that used by Alonzo and Pepe \cite{alonzo2002distribution} for ROC regression. Earlier work by Pepe \cite{pepe2000interpretation} noted that if we define $U_{ij} = \mathbbm{1}(B_{1i}(\lambda) > B_{0j}(\lambda))$ where $i$ and $j$ index treated and untreated patients, then $S_{1|x}(S_{0|x}^{-1}(\omega_j))$ is equivalent to the expectation of $U_{ij}$ conditional on $S_{0|x}(B_{0j}) = \omega_j$ and $X=x$. Given estimated quantiles $\widehat{\omega}_j$, standard GLM procedures can be used to estimate $S_{1|x}(S_{0|x}^{-1}(\omega))$. Because $U_{ij}$ must be defined for every pair of treated and untreated units, this approach is intractable for large samples. Instead, we use an alternative approach proposed by Alonzo and Pepe \cite{alonzo2002distribution} which improves on the efficiency of the original estimator. Let $\Omega$ be a finite set of values between $0$ and $1$. For $\omega \in \Omega$, if we define $U_{i\omega} = \mathbbm{1}(B_{1i}(\lambda) > S_{0|X_i}^{-1}(\omega))$ for each unit $i$ receiving treatment, then $\mathbb{E}[U_{i\omega}] = S_{1|X_i}(S^{-1}_{0|X_i}(\omega))$. For a given probability, $\omega \in \Omega$, this definition of $U_{i\omega}$ gives the indicator of whether a treated unit's observed INB is greater than the $\omega$-quantile of the distribution of INBs in an untreated population. In practice, because $S^{-1}_{0|X}(\omega)$ is unknown, $U_{i\omega}$ is defined using a consistent estimator of the quantile function $U_{i\omega} = \mathbbm{1}(B_{i}(\lambda) > \widehat{S}_{0|X_i}^{-1}(\omega))$. Alonzo and Pepe \cite{alonzo2002distribution} suggest using empirical estimates of the quantiles when possible (i.e. for discrete $X$) and quantile regression \cite{koenker1978regression} otherwise. 
\par The conditional expectation of $U_{i\omega}$ can be modeled using any flexible regression approach. One sensible modeling choice is the probit form: $\mathbb{E}[U_{\omega}|X] = \Phi(\beta_0 + \beta_1 X + \beta_2 \Phi^{-1}(\omega))$ where $\Phi(\cdot)$ is the cumulative distribution function of the standard normal distribution. This is the form that would arise if $B_1(\lambda)$ and $B_0(\lambda)$ were normally distributed or if they could be monotonically rescaled to be normally distributed. Hanley \cite{hanley1988robustness} shows that when fitting ROC curves using the binormal form, estimates of AUC exhibit low bias even when the model is misspecified. Because we characterize NBS as the area under the fit binormal model, it can be seen as analogous to AUC. Once the conditional expectation has been modeled, NBS can be estimated through numerical integration using standard software \cite{piessens2012quadpack}. Under the probit model, our estimate is given by
\begin{equation}\label{eqn:regest}
    \widehat{\theta}(\lambda|X) = \int_0^1 \Phi(\widehat{\beta}_0 + \widehat{\beta}_1 X + \widehat{\beta}_2 \Phi^{-1}(\omega)) d\omega.
\end{equation}
\par Testing whether $X$ influences $\theta(\lambda|X)$ is equivalent to testing whether the coefficient(s) associated with $X$ are equal to zero. For example, under the probit model this is a test of if $\beta_1 = 0$. Because multiple values $U_{i\omega}$ are defined for each individual (one for each $\omega \in \Omega$), model-based variance estimates that assume independent observations are invalid. Instead, we propose the nonparametric bootstrap hypothesis test. An important consideration in the regression of NBS is the choice of $\Omega$. For sets consisting of $N_\omega$ equally spaced points, $\Omega = \{j/(N_\omega+1); j = 1,..., N_\omega \}$, selecting $N_\omega = N_0-1$ is equivalent to the estimator originally proposed by Pepe \cite{pepe2000interpretation} and is maximally efficient. However, Alonzo and Pepe \cite{alonzo2002distribution} showed that smaller choices of $N_\omega$ were able to achieve close to maximal efficiency while mitigating computational burden.

\subsection{Adjustment for measured confounding}\label{sec:standardization}

\par Suppose that there are measured covariates, $L$, which affect the probability of receiving treatment as well as the cost or effectiveness outcomes. Our proposed estimation procedure consists of two steps: a Monte Carlo standardization step, and a regression step. The first step constructs two pseudo-populations which differ in treatment status, but where the distribution of confounding variables, $L$, is equivalent. Let $\widehat{f}(z,y|A,X,L)$ denote some model for the conditional distribution of cost and effectiveness measures. The pseudo-population under treatment $A=a$ is built by drawing $M$ observations from the empirical distribution of the baseline covariates, $\widehat{F}(L,X)$, and then using the sampled values of $L$ and $X$ to sample from the fit model $\widehat{f}(z,y|A=a, X, L)$. From these pseudo-populations, we may sample from the joint distributions of the INBs and $X$ among treated and untreated units, $\left(B_1(\lambda), X\right)$ and $\left(B_0(\lambda), X\right)$ respectively. This methodology ensures that INBs are sampled from populations which receive different treatments, but have the same underlying distribution of measured confounders. This procedure is semiparametric because, while parametric models may be used to fit the distributions of $Z$ and $Y$, the joint distribution $F(X,L)$ is estimated nonparametrically. The Monte Carlo standardization procedure is summarized below:
\begin{enumerate}
    \item Model the conditional distributions $f(Z|A,X,L)$ and $f(Y|A,X,L,Z)$ where $f(\cdot)$ and $F(\cdot)$ denote the probability density function and the cumulative density function of a random variable. Let $\widehat{f}(Z|A,X,L)$ and $\widehat{f}(Y|A,X,L,Z)$ denote the estimated distributions.
    \item For $a=0,1$:
    \begin{itemize}
        \item Sample $\widetilde{l}_m$ and $\widetilde{x}_m$ for $m=1,..., M$ from the empirical distribution $\widehat{F}(L,X)$.
        \item For $m=1,..., M$ sample $\widetilde{z}_m$ and $\widetilde{y}_m$ from $\widehat{f}(Z|A=a,\widetilde{x}_m,\widetilde{l}_m)$ and $\widehat{f}(Y|A=a, \widetilde{x}_m, \widetilde{l}_m, \widetilde{z}_m)$
        \item Calculate $\widetilde{B}_{a,m}(\lambda)$ for each pair $(\widetilde{z}_m, \widetilde{y}_m)$
    \end{itemize}
\end{enumerate}
The models for $f(Z|A,X,L)$ and $f(Y|A,X,L,Z)$ can be flexibly estimated to reduce bias. For example, zero-inflated gamma regression models can be used to accommodate skewed cost distributions with structural zeros. By including effectiveness as a predictor of cost, this procedure allows for correlation between $Z$ and $Y$.
\par The second step of our estimation procedure entails regressing the NBS using the sampled INBs. Because $(B_1(\lambda), X)$ and $(B_0(\lambda), X)$ are sampled from populations where treatment is independent of measured confounders, NBS regression proceeds as described in Section \ref{sec:covarnbs}. The regression step is outlined below: 
\begin{enumerate}
    \item For each $\omega \in \Omega$, use sampled values $\widetilde{B}_{0,m}(\lambda)$ to estimate $S_{0|X}^{-1}(\omega)$.
    \item For each generated $\widetilde{B}_{1,m}(\lambda)$ and $\omega \in \Omega$, define $U_{m\omega} = \mathbbm{1}(\widetilde{B}_{1,m}(\lambda) > \widehat{S}_{0|X}^{-1}(\omega))$.
    \item Fit a model for $\mathbb{E}[U_{m\omega}|X]$
    \item For $X=x$, estimate the subgroup-specific NBS by numerically integrating:
    \begin{equation*}
        \widehat{\theta}(\lambda|X=x) = \int_0^1 \widehat{\mathbb{E}}[U_{i\omega}|X=x]
    \end{equation*}
\end{enumerate}

\subsection{Informative Censoring}\label{sec:ipcw}

\par Up until this point, we have developed methods which are suitable for cases with no censoring, e.g. when $Z$ is a continuous outcome with no missing data. Often, the effectiveness outcome of greatest clinical interest is time to death post treatment. Because patients followed over time may be lost to follow-up, we introduce methodology to account for censoring when estimating NBS. Suppose we are interested in estimating costs up until a maximum time, $\tau$. For observation $i$, let $C_i$ be the censoring time, $T_i$ be the survival time, and $T_i^* = \text{min}(T_i, \tau)$. Define $\delta_i = \mathbbm{1}(C_i < T_i)$ and $\delta_i^* = \mathbbm{1}(C_i < T_i^*)$ be indicators of censored survival time and censored costs. The observable data are $\delta$, $\delta^*$, and $Z = min(T_i, C_i)$. Because costs at time of censoring can be correlated with costs at event time, it is necessary to use IPCW to estimate $f(Y|A,X,L,Z)$. Let $G(t) = P(t \le C)$ denote the probability that the censoring time is beyond $t$. If we assume $T$ and $C$ are independent, then $G(t)$ can be modeled within each treatment group using the Kaplan-Meier product limit estimator based on the data $(C, \delta^*)$. If instead we assume that survival and censoring times are independent only once we condition on treatment status and other measured variables, $W$, then we can use the stratified cox model to estimate $G(t)$. Under this model, we assume that the hazard function for censoring is given by:

\begin{equation*}
h(t|V,W) = \text{exp}(\eta^T W)h_V(t)
\end{equation*}

where V denotes stratifying variables (treatment status, $A$, and other discrete covariates). The coefficients in this model, $\eta$, can be estimated using the Cox partial likelihood \cite{cox1975partial}:

\begin{equation*}
    \widehat{G}(T_i^*|V_i,W_i) = \text{exp}\left(-\sum_{j=1}^{n}\frac{\delta_j \mathbbm{1}(V_i = V_j, T_i^* > X_j)\text{exp}(\widehat{\eta}^T W_i)}{\sum_{k=1}^n \mathbbm{1}(V_i = V_k, X_k \ge X_j)\text{exp}(\widehat{\eta}^T W_k)} \right)
\end{equation*} 

For each individual with uncensored cost data, IPCW is defined as

\begin{equation*}
    w_i = \frac{\delta_i^*}{\widehat{G}(T_i^*)}
\end{equation*}

The weights, $w_i$, can be incorporated into the Monte Carlo procedure when modeling $f(Y|A,X,L,Z)$. For example, suppose we assume that costs follow a log-normal distribution. To fit $f(Y|A,X,L,Z)$, we can perform weighted log-normal regression using only patients with fully observed cost data (i.e. those with $\delta_i^* = 1$). Any modeling procedure that incorporates weighting may be used to estimate the cost distribution. The resulting estimate, $\widehat{f}(Y|A,X,L,Z)$, can be used in the standardization procedure described in Section \ref{sec:standardization}.

\section{Simulations}\label{sec:simulations}

\par We perform a simulation study to determine the finite-sample properties of the proposed regression estimator. Although data used for cost-effectiveness analyses often have thousands of observations, it is beneficial to understand how our methodology performs in both small and large sample settings. We consider simulation settings with low sample size ($n = 500$) and high sample size ($n=5000$) as well as low ($10\%$), medium ($30\%$), and high ($50\%$) rates of censoring. Finally, we consider the case where the stratifying variable is and is not associated with $\theta(\lambda|X)$. For each simulation, we estimate the NBS under two WTP thresholds, $\lambda = 2$ and $12$. These represent the cost a payer is willing to incur for a unit change in effectiveness.
\par We generate the data as follows: $L \sim N(0,1)$, $X \sim \text{Bernoulli}(p_x =  0.25)$, $A \sim \text{Bernoulli}(p_a = \text{expit}(L))$, $T \sim \text{Weibull}(k=2, \lambda =  \text{exp}(4.05 + 0.15 A + 0.2 L + \beta_{x} X + \beta_{ax} A\times X))$, $C \sim \text{Weibull}(k=2, \lambda =  \text{exp}(\gamma + 0.5 A))$, and $Y \sim \text{Lognormal}(\mu = 4.2 + 0.002T + 0.5A, \sigma^2 = 0.16)$. In simulations where $X$ is associated with $\theta(\lambda|X)$ we set $\beta_x = 0.1$ and $\beta_{ax} = 0.5$. Under these parameter values $\theta(2|X=0) = 0.353$, $\theta(2|X=1) = 0.588$, $\theta(12|X=0) = 0.527$ and $\theta(12|X=1) = 0.746$. 
\par Treated subjects are expected to have greater costs and survival times than those which are untreated. Patients with $X = 1$ benefit more from treatment (with respect to survival) than those with $X = 0$. The separation between the distribution of INBs between treated and untreated subjects is larger in a population where $X=1$ than in one with $X=0$. In the setting where $\theta(\lambda|X)$ does not depend on $X$, both $\beta_x$ and $\beta_{ax}$ are set to zero. In these simulations, $\theta(2|X) = 0.353$ and $\theta(12|X) = 0.527$ regardless of the value of $X$. In simulations where $X$ affects NBS, we set $\gamma$ to $0.5119$, $0.4410$, and $3.9600$ to induce $10\%$, $30\%$, and $50\%$ censoring. When $X$ does not affect NBS, these values are $5.007$, $4.315$, and $3.8760$, respectively.
\par To estimate $\theta(\lambda|X)$, we use the standardization procedure described in Section \ref{sec:covarnbs}. A Weibull regression model is used to estimate the distribution of $T$ and a log-normal model is used for costs. Because $Y$ is informatively censored, we use IPCW when estimating the cost model. To estimate the probability of being uncensored, we use the Cox proportional hazards model. In the standardization step, we draw $M=5000$ observations within each level of $A$. For the regression of $S_{1|X}(S^{-1}_{0|X}(\omega))$, we obtain $\widehat{S}^{-1}_{0|X}(\omega))$ for $\omega\in \Omega$ using the empirical quantiles within levels of $X$. The previously described probit link is used to model $U_{i\omega}$ conditional on $X$. Standard errors are estimated as the standard deviation of $B = 300$ bootstrap replicates. The nominal type-I error rate is set to $0.05$ for our hypothesis tests. Hypothesis tests are performed using symmetric $95\%$ CIs based on the bootstrap replicates.
\par For each setting, we simulate $1000$ datasets, estimate the NBS conditional on $X$, and perform the nonparametric bootstrap hypothesis test. We report the mean estimate of the NBS across simulations, the mean bootstrap estimated standard error of the NBS, the standard deviation of estimated NBS across simulations, and the proportion of hypothesis tests in which the null hypothesis was rejected. Tables \ref{tab:sim-eff} and \ref{tab:sim-noeff} provide simulation results for the settings where $X$ affects and does not affect the NBS, respectively.
\par Average point estimates of NBS are close to their true values for all tested values of $X$ and $\lambda$. Bootstrap estimated standard errors closely approximate the empirical standard errors observed accross simulations. Due to a lower effective sample size, standard errors increase as the proportion of censored units increases. Similarly, higher variability in the estimates of $\theta(\lambda|X=0)$ can be attributed to the fact that there is a lower prevalence of $X=0$ in our simulations. In simulations where $X$ affects NBS, power is lowest when there is a sample size of $500$, $50\%$ censoring, and a WTP of $2$. All other observed rejection rates exceed $0.80$. When $X$ does not affect cost-effectiveness, the probability of rejecting the null is close to the desired $0.05$. All simulations are performed in R 3.6.0. Code to replicate simulation results and for the following data example are provided in our supplemental materials.

\subsection{Sensitivity to unmeasured confounding}

\par In settings where we estimate NBS using data from observational databases, we must consider the potential effects of unmeasured confounding. Because the Monte Carlo procedure described in Section \ref{sec:standardization} only ensures that INBs are sampled from populations with the same underlying distribution of measured covariates, it is necessary to understand how estimates of NBS may be affected by unmeasured confounding. We perform simulations to determine how sensitive the proposed regression estimator of NBS is to unmeasured confounding. We simulate two standard normal variates, $U_1$ and $U_2$. The confounder $U_1$ influences treatment probability and survival time, while $U_2$ affects treatment and cost. We consider two settings, (1) where only the exposure-survival relationship is confounded and (2) where only the exposure-cost relationship is confounded. Variables $L$, $X$, and $C$ are simulated as in previous simulations, $A\sim\text{Bernoulli}\left(\text{expit}(L +  \gamma_1 U_1 + \gamma_2 U_2)\right)$, $T \sim \text{Weibull}(k=2, \lambda =  \text{exp}(4.05 + 0.15 A + 0.2 L - \eta_1 U_1 +  0.1 X + 0.5 A\times X))$, and $Y \sim \text{Lognormal}(\mu = 4.2 + 0.002T + 0.5A + \eta_2 U_2, \sigma^2 = 0.16)$. The parameters $\gamma_1$, $\gamma_2$, $\eta_1$ and $\eta_2$ control the strength of unmeasured confounding. In simulations where the exposure-survival relationship is confounded, we examine the effect of low ($\gamma_1 = 0.5$, $\eta_1 = 0.05$), medium ($\gamma_1 = 0.75$, $\eta_1 = 0.15$), and high ($\gamma_1 = 1$, $\eta_1 = 0.3$) levels of confounding. The parameters $\gamma_2$ and $\eta_2$ equal zero in this setting. When the exposure-cost relationship is confounded, $\gamma_1 = \eta_1 = 0$. Low, medium, and high levels of confounding are simulated by setting ($\gamma_2 = 0.5$, $\eta_2 = 0.17$), ($\gamma_2 = 0.75$, $\eta_2 = 0.5$), and ($\gamma_2 = 1$, $\eta_2 = 1$) respectively
\par We simulate $1000$ datasets with $5000$ observations and a $30\%$ censoring rate. We do not account for $U_1$ or $U_2$ when estimating NBS using the Monte Carlo standardization procedure. Simulation results for these simulations are provided in Table \ref{tab:cnbs-sensitivity}. As the level of confounding increases, the mean estimate of the NBS becomes more biased. When confounding is strongest, percent bias exceeds $10\%$. Because cost is given less weight in INB for large WTP, the bias from unmeasured exposure-cost confounding decreases as WTP increases. Similarly, bias increases with WTP when there is unmeasured confounding of the exposure-survival relationship. Bootstrap estimated standard errors are able to capture the true variability in estimates of NBS regardless of the strength of confounding.

\section{Endometrial Cancer Study}\label{sec:data}

\par The standard treatment for early-stage endometrial cancer is complete hysterectomy. Patients sometimes receive additional adjuvant radiation or chemotherapy to decrease the risk of recurrence \cite{latif2014adjuvant}. However, there is insufficient evidence that adjuvant radiation therapy increases overall survival \cite{shaeffer2005adjuvant}. Because more than half of the cost associated with the treatment of endometrial cancer is accrued in the period after initial treatment \cite{mariotto2011projections}, it is important to ensure that only those patients who are likely to benefit from adjuvant therapies receive additional treatment. We illustrate how to determine the cost-effectiveness of adjuvant radiation and chemotherapy using empirically-based synthetic data developed by Spieker et al. \cite{spieker2019net}. This dataset was developed based on data from a large observational endometrial cancer database. Followup for patients in the original database is insufficient to perform a meaningful cost-effectiveness analysis. Covariates in the synthetic data are drawn from their empirical distribution in the cancer database. Cost and survival outcomes are simulated based on relations observed in the database and from literature surrounding the prognoses of stage I and stage II endometrial cancer \cite{spieker2020causal, shaeffer2005adjuvant, susumu2008randomized}. The data have covariate, treatment, and outcome distributions similar to what would be observed in practice. The use of synthetic data is motivated by a desire to make the code and data for this analysis available as well as to ensure sufficient followup. 
\par We consider three possible courses of treatment post-hysterectomy: (1) receipt of adjuvant radiation therapy (RT), (2) receipt of adjuvant chemotherapy (CT), and (3) receipt of neither therapy, hereafter referred to as control. In our first analysis we compare those receiving RT to those receiving control, and in our second we compare CT to control. Because the data are simulated, the results of our analysis are to be used solely for illustrating how NBS regression can be used in a practical setting. This work does not make use of data from human subjects and does not require IRB approval. 
\par The data contain monthly follow-up information on $N = 13526$ endometrial cancer patients. The upper bound on follow-up time is ten years, at which point all observations are administratively censored. Our goal is to evaluate the association between two measures of patient health, cancer stage and Charlson comorbidity index \cite{charlson1994validation}, and the cost-effectiveness of adjuvant RT or CT compared to control. We define cost-effectiveness using medical costs in USD(\$) and survival time in years. We assume that age at diagnosis, cancer stage, Charlson index, baseline receipt of RT or CT, and number of hospitalizations in the first-month post-surgery are confounders of the exposure-outcome relationship. Treatment status is defined over the period from two-to-four months post-hysterectomy. For example, if a patient received adjuvant RT during this period, then they are included in the RT group in our analysis regardless of their treatment status throughout the remainder of the study period. If a patient received both RT and CT during this period, they are included in both treatment groups. The mean age at diagnosis is $73.70$ (SD = $6.59$), and most patients have stage I cancer ($93.6\%$). Charlson comorbidity indices are between zero and five, with $51.32\%$ of patients having an index of zero. In the first month post-hysterectomy, the baseline period, $74$ ($0.55\%$) patients received RT and $810$ ($5.99\%$) received CT. From two-to-four months post-surgery, these figures were $3440$ ($25.43\%$) and $305$ ($2.25\%$), respectively.

\subsection{Analysis}

\par We wish to estimate NBS comparing each adjuvant therapy to control conditional on cancer stage and Charlson comorbidity index. We consider a range of WTP (from $\$50k$ to $\$120k$ per year),cancer stages (stages I and II), and categorized Charlson indices ($0$, $1$, and $2+$). Because baseline covariates affect both treatment status and patient outcomes, we apply the standardization procedure described in Section \ref{sec:standardization}. We assume survival time can be modelled using a Weibull distribution. To account for structural zeros in cost data, we use a two-stage zero-inflated cost model where the probability of having zero costs is estimated using a logistic model and the distribution of nonzero costs is fit using a log-normal model. All models are fit conditional on treatment status and confounders. For each treatment, we draw $M = 10000$ Monte Carlo observations from the fit models for survival and medical costs and use a probit model when finding the regression estimator of the NBS. For our analysis of RT versus control, we model the NBS using the form:
\begin{align*}
    \theta(\lambda|\text{Charlson, Stage}) = \int_{0}^1 \Phi\bigg(\widehat{\beta}_{0}^R \ + \ &\widehat{\beta}_{1}^R \ \text{Stage II} \ + \ \widehat{\beta}_{2}^R \ \text{Charlson(1)} \\ + \ &\widehat{\beta}_{3}^R \ \text{Charlson(2+)} \ + \widehat{\beta}_{4}^R \Phi^{-1}(\omega)\bigg) d\omega
\end{align*}
The variables $\text{Charlson(1)}$ and $\text{Charlson(2+)}$ are indicators of whether a patient's Charlson score is equal to one or at least two, respectively. Analogous models are used when estimating the NBS comparing CT to control conditional on cancer stage and Charlson index. Parameters for the regression model comparing CT to control are denoted by the superscript $C$. Hypothesis testing is based on $B = 1000$ nonparametric bootstrap replicates. 

\subsection{Results}

\par
\begin{figure}
\centering
\includegraphics[width = 0.9\linewidth]{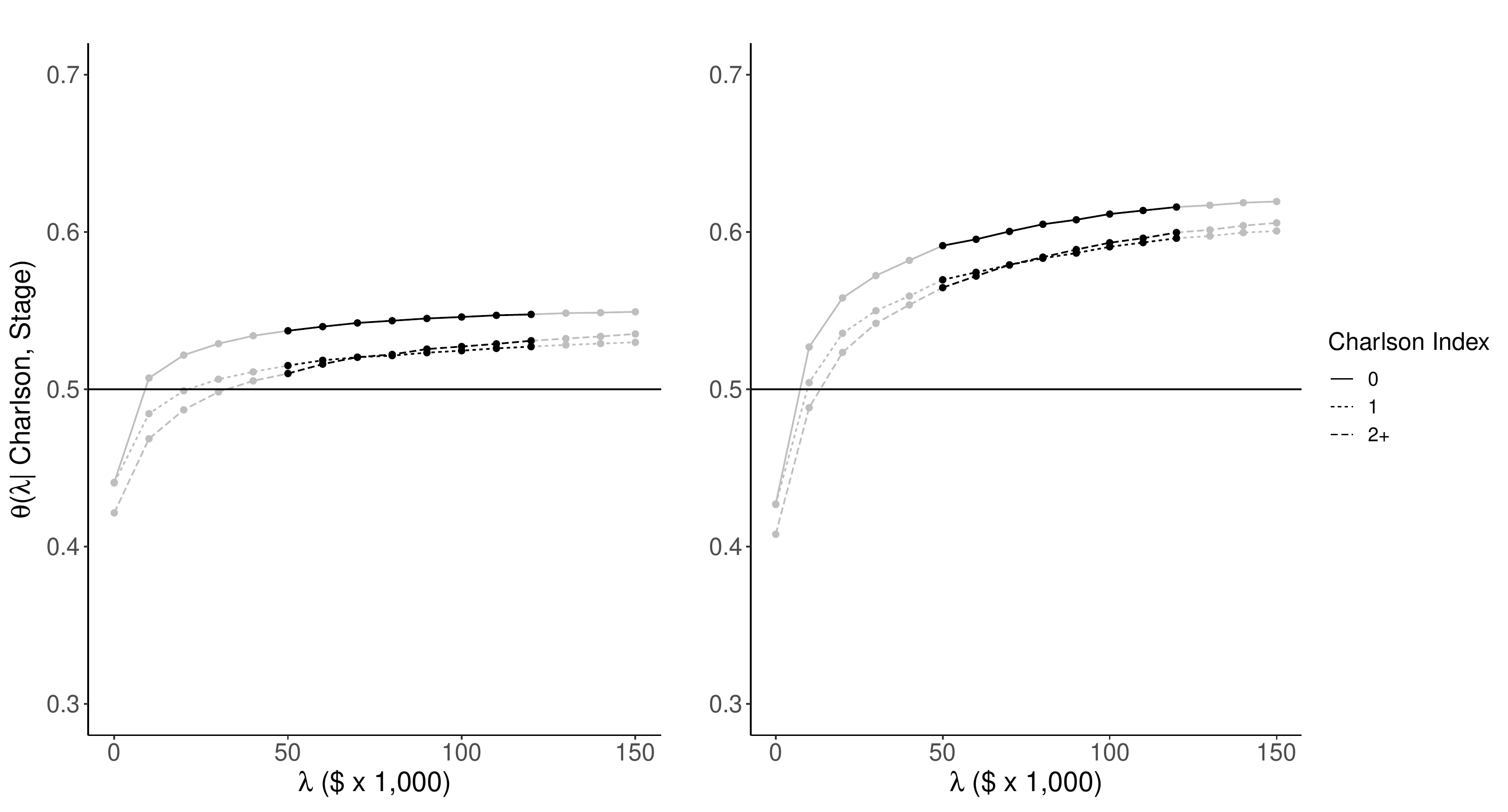}
\caption{Estimates of the NBS comparing RT to control as a function of the WTP, $\lambda$. CED curves are provided for patient cohorts conditional on cancer stage and Charlson comorbidity index. Results for patients with stage I cancer are provided on the left, while those for patients with stage II cancer are on the right. Estimated NBS within the range of primary interest are denoted in black. Gray indicates estimated NBS outside of this region, provided to observe the behavior of the NBS.}
\label{fig:ced-rt}
\end{figure}

\begin{figure}
\centering
\includegraphics[width = 0.9\linewidth]{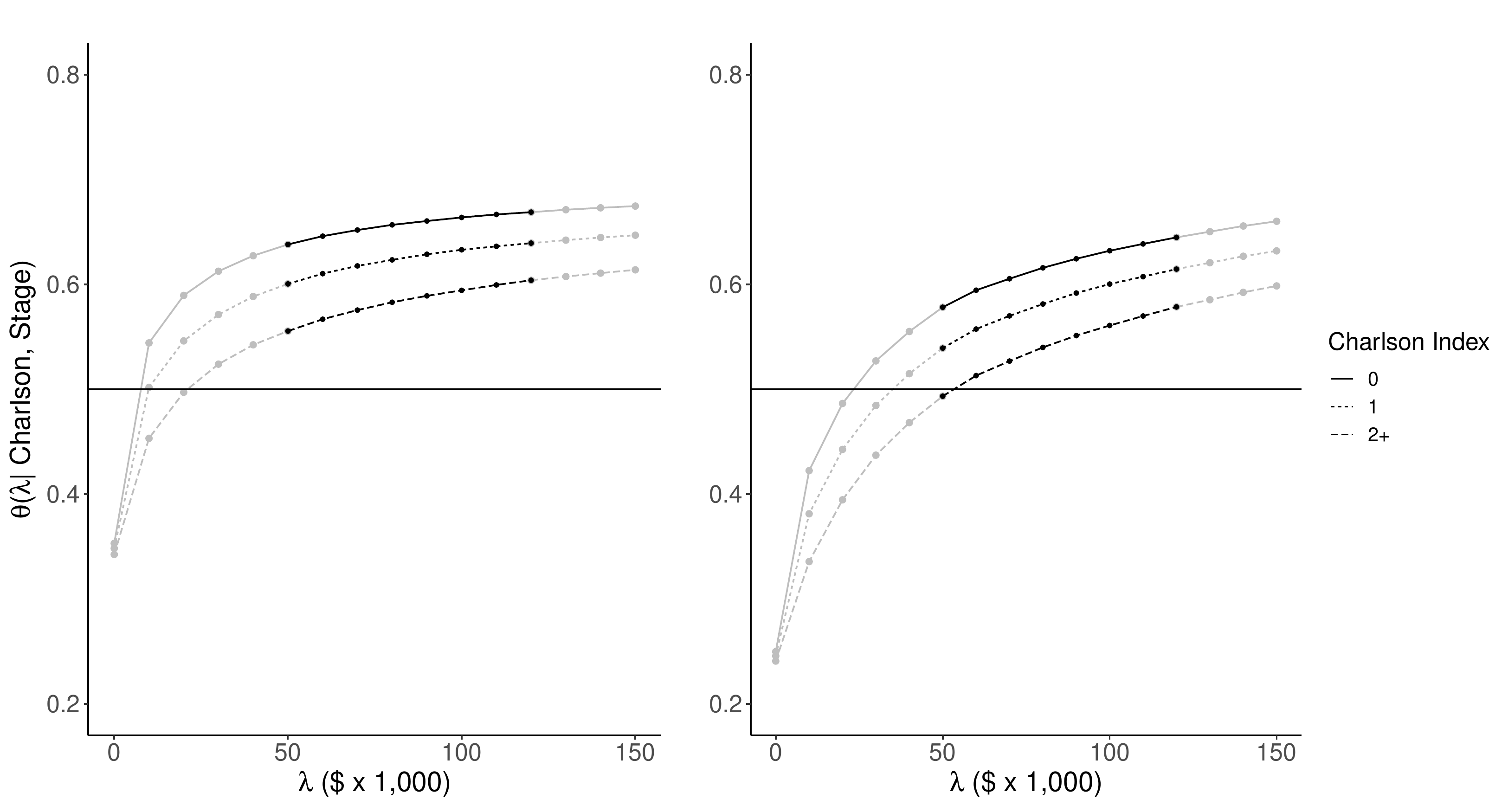}
\caption{Estimates of the NBS comparing CT to control as a function of the WTP, $\lambda$. CED curves are provided for patient cohorts conditional on cancer stage and Charlson comorbidity index. Results for patients with stage I cancer are provided on the left, while those for patients with stage II cancer are on the right. Estimated NBS within the range of primary interest are denoted in black. Gray indicates estimated NBS outside of this region, provided to observe the behavior of the NBS.}
\label{fig:ced-ct}
\end{figure}

 In Table \ref{tab:nbs_charlson_stage}, we provide estimates of the NBS for each of the treatments as a function of Charlson index and cancer stage. In almost every setting, the estimated NBS is greater than $0.5$, indicating that patients are likely to benefit from treatment regardless of their cancer stage or number of comorbidities at diagnosis. Because there are few subjects who have stage II cancer and receive adjuvant chemotherapy ($n = 45$), estimates of the NBS within this cohort too uncertain to claim cost-effectiveness. For WTP of $\$50k$, the estimated NBS comparing RT to control among patients with stage I cancer and a comorbidity index of zero is $0.54$ with $95\%$ CI $(0.51, 0.55)$. We conclude that a randomly selected patient receiving RT will have a greater INB than a patient receiving control $54\%$ of the time. Adjuvant radiation is cost-effective for this patient cohort. The $95\%$ CI for $\theta(\lambda=50|\text{Stage I, Charlson(2+)})$ comparing RT to control is $(0.48, 0.53)$. Thus, there is inconclusive evidence on the cost-effectiveness of RT compared to control for patients with stage I cancer and multiple comorbidities. Figures \ref{fig:ced-rt} and \ref{fig:ced-ct} provide cost-effectiveness determination curves (CED) comparing adjuvant radiation to control and adjuvant chemotherapy to control, respectively. The CED is defined by Spieker et al. \cite{spieker2019net} as a graphical tool which shows how NBS is affected by changes in WTP. The bold segment of the plots represent estimated NBS at WTP between $\$50\text{k}$ and $\$120\text{k}$, the range of primary interest \cite{shiroiwa2010international}. Estimates of NBS outside of this range are in gray to illustrate the behavior of NBS at extreme WTP.
\par For each treatment and WTP, we conduct a hypothesis test to evaluate if there is an association between either cancer stage or Charlson comorbidity index and NBS. As per Section \ref{sec:covarnbs}, we can determine whether cancer stage is associated with NBS after controlling for comorbidities by testing the null hypotheses $H_{0}^{R}: \beta_{1}^{R} = 0$ and $H_{0}^{C}: \beta_{1}^{C} = 0$ against their alternatives. At $\lambda = \$50\text{k}$, the p-values for these tests are $0.019$ and $0.366$, respectively. There is evidence that, after adjusting for Charlson comorbidity indices, the cost-effectiveness of adjuvant radiation may depend on cancer stage. However, there is insufficient evidence to say same for adjuvant chemotherapy at this WTP. At a WTP of $\$120\text{k}$, the p-values for the tests of the same hypotheses are $0.007$ and $0.543$. Again, there is evidence that cancer stage is associated with the cost-effectiveness of adjuvant radiation therapy but not of chemotherapy. Among patients with a Charlson comorbidity score of zero, the estimated NBS at a WTP of $\$120\text{k}$ is $0.55$ for patients with stage I cancer and $0.62$ for those with stage II cancer. This indicates that patients with more advanced cancers may benefit more from receiving adjuvant radiation therapy than those with stage I cancer. A similar conclusion is drawn at a WTP of $\$50\text{k}$.
\par To test whether patient's comorbidity indices are associated with NBS after accounting for cancer stage, we perform a test of the hypotheses $H_0^R: \beta_2^R = \beta_3^R = 0$ and $H_0^C: \beta_2^C = \beta_3^C = 0$ against their respective alternatives. At a WTP of $\$50\text{k}$, the p-value for the test of parameters in the adjuvant radiation model is $ 0.022$, while that for chemotherapy less than $0.005$. At a WTP of $\$120\text{k}$, the test for association between the Charlson comorbidity index and the cost-effectiveness of adjuvant radiation results in a p-value of $0.076$. At the same WTP the test for adjuvant chemotherapy, again, has a p-value less than $0.005$. At both considered WTP, there is evidence that Charlson comorbidity score can help determine which patients may most benefit from adjuvant chemotherapy. However, while there is evidence that the cost-effectiveness of adjuvant radiation is affected by Charlson comorbidity index at a WTP of $\$50\text{k}$, there is insufficient evidence at the larger WTP. Because medical costs are given greater weight for lower WTP, this suggests that comorbidities may have a greater effect on medical costs than on patient survival. For patients being treated with adjuvant chemotherapy, point estimates of the NBS in Table \ref{tab:nbs_charlson_stage} indicate that patients with greater Charlson comorbidity indices may not receive as much benefit as those with fewer comorbidities, regardless of cancer stage.

\section{Discussion}

\par In this paper, we introduce a regression framework for NBS that allows us to explain variability in cost-effectiveness arising from covariates. The proposed method is the first to enable estimation of NBS within levels of measured covariates and allow for testing the effects of covariates on NBS. Regression of NBS is done in three steps: (1) estimate the distribution of cost and effectiveness outcomes conditional on treatment and measured covariates, (2) implement Monte Carlo standardization to sample INBs from treated and untreated populations with the same underlying distribution of confounders, and (3) apply ROC regression techniques developed by Alonzo and Pepe \cite{alonzo2002distribution} to estimate NBS. Our proposed methodology generalizes the work of Spieker et al. \cite{spieker2019net} on estimating NBS to allow for covariate adjustment and hypothesis testing. Understanding how patient characteristics influence cost-effectiveness can help policy makers allocate resources towards groups which are most likely to benefit from treatment. 
\par In cost-effectiveness analyses where patients are lost to follow-up, estimators that assume independence between costs at time of censoring and those at time of death are known to be inconsistent \cite{lin1997estimating}. We provide methodology for incorporating IPCW in models for the conditional distribution of treatment costs. When the probability of censoring depends on measured covariates, we describe how Cox proportional hazards models can be used to determine covariate-specific IPCW. In simulations, we find that our method is able to attain low bias and adequate power in settings with small sample size and up to $50\%$ censoring. Moreover, estimates of standard error obtained using the nonparametric bootstrap are representative of empirical standard errors across various levels of censoring. 
\par Our proposed estimation procedure assumes that models for the conditional distributions of cost and effectiveness are properly specified. Flexible modeling techniques present a potential route to reduce the risk of model misspecification. Bayesian nonparametric methods for zero-inflated data have been developed by Oganisian et al. \cite{oganisian2018bayesian} to estimate cost distributions, but extensions of this methodology to cost-effectiveness analyses have not been explored. Similarly, ensemble learning algorithms such as Super Learner \cite{van2007super} could be used to model complex cost and effectiveness distributions. For example, in their doubly robust estimator of NMB, Li et al. \cite{li2018doubly} successfully use Super Learner to estimate both propensity score and outcome models. In future work, it would be useful to determine the best ways to incorporate these methods into regression of NBS. After the standardization step, parametric regression models for NBS may also be misspecified. Prior work by Hanley \cite{hanley1988robustness} has examined the robustness of the binormal form for modeling ROC curves. Their results suggest that estimates of AUC arising from the binormal form, and analogously NBS, may exhibit low bias even when the form is misspecified. Additionally, because the underlying distribution of INBs are left unspecified, the proposed NBS regression framework can be described as “distribution-free” \cite{alonzo2002distribution}. 
\par Because cost-effectiveness data are often drawn from observational databases, unmeasured confounding may influence estimates of NBS.  In simulations with low levels of unmeasured confounding, our regression estimator of NBS exhibits small to moderate bias (between $2\%$ and $8\%$ bias). As expected, we found that as the strength of confounding increases, estimates of NBS become more biased. In any studies involving observational databases, researchers should consider potential sources of unmeasured confounding. Methods to assess sensitivity of NBS to unmeasured confounding may also be useful and will be a subject of future work.
\par Note that NBS is defined in terms of baseline treatment status. In our endometrial cancer example, the comparison groups are those who received some adjuvant therapy in the period from two to four months after surgery and those who did not. Because treatment effects may depend on the duration or frequency of treatment, conclusions concerning cost-effectiveness may change depending on whether treatment is considered time-stable, as in our definition, or time-dependent. To estimate cumulative medical costs under a time-dependent treatment strategy, Spieker et al. \cite{spieker2018analyzing} developed a nested g-computation procedure. Similar methodologies that can incorporate time-dependent treatment may be useful when we are interested in comparing the cost-effectiveness of multiple treatment strategies and are the subject of future work.
\par Because patient covariates may influence the cost-effectiveness of a treatment, our proposed regression framework for NBS provides a useful approach to identify which groups may benefit the most from treatment.  The results from our synthetic data analysis illustrate how we can evaluate the effects of cancer stage or number of comorbidities on the cost-effectiveness of adjuvant therapies for endometrial cancer patients. This information may be useful to policy makers aiming to better allocate resources.

\bibliographystyle{apacite}
\bibliography{sample}

\clearpage

\begin{table}
    \centering
    \begin{tabular}{ccccccccc}
    \toprule
        Cens. & Sample Size & $\lambda$ & X &  $\theta(\lambda|X)$ & Est. & $\widehat{\text{SE}}$ & ESE & Pr. Reject  \\
    \midrule
    \midrule
        10\% & 500 & 2 & 0 & 0.353 & 0.352 & 0.028 & 0.030 & 0.986 \\
        10\% & 500 & 2 & 1 & 0.588 & 0.589 & 0.049 & 0.049 & $-$ \\
        10\% & 500 & 12 & 0 & 0.527 & 0.526 & 0.028 & 0.030 & 0.988 \\
        10\% & 500 & 12 & 1 & 0.746 & 0.745 & 0.038 & 0.038 & $-$ \\
        10\% & 5000 & 2 & 0 & 0.353 & 0.354 & 0.011 & 0.010 & 1.000 \\
        10\% & 5000 & 2 & 1 & 0.588 & 0.587 & 0.019 & 0.020 & $-$ \\
        10\% & 5000 & 12 & 0 & 0.527 & 0.527 & 0.011 & 0.010 & 1.000  \\
        10\% & 5000 & 12 & 1 & 0.746 & 0.745 & 0.015 & 0.015 & $-$ \\
        30\% & 500 & 2 & 0 & 0.353 & 0.351 & 0.032 & 0.032 & 0.924 \\
        30\% & 500 & 2 & 1 & 0.588 & 0.591 & 0.061 & 0.065 & $-$ \\
        30\% & 500 & 12 & 0 & 0.527 & 0.526 & 0.032 & 0.032 & 0.964 \\
        30\% & 500 & 12 & 1 & 0.746 & 0.746 & 0.045 & 0.046 & $-$ \\
        30\% & 5000 & 2 & 0 & 0.353 & 0.352 & 0.012 & 0.012 & 1.000 \\
        30\% & 5000 & 2 & 1 & 0.588 & 0.588 & 0.023 & 0.023 & $-$ \\
        30\% & 5000 & 12 & 0 & 0.527 & 0.526 & 0.012 & 0.012 & 1.000 \\
        30\% & 5000 & 12 & 1 & 0.746 & 0.745 & 0.017 & 0.017 & $-$ \\
        50\% & 500 & 2 & 0 & 0.353 & 0.353 & 0.041 & 0.040 & 0.694 \\
        50\% & 500 & 2 & 1 & 0.588 & 0.580 & 0.083 & 0.087 & $-$ \\
        50\% & 500 & 12 & 0 & 0.527 & 0.527 & 0.037 & 0.036 & 0.842 \\
        50\% & 500 & 12 & 1 & 0.746 & 0.742 & 0.057 & 0.055 & $-$ \\
        50\% & 5000 & 2 & 0 & 0.353 & 0.353 & 0.015 & 0.016 & 1.000 \\
        50\% & 5000 & 2 & 1 & 0.588 & 0.586 & 0.034 & 0.033 & $-$ \\
        50\% & 5000 & 12 & 0 & 0.527 & 0.527 & 0.013 & 0.014 & 1.000 \\
        50\% & 5000 & 12 & 1 & 0.746 & 0.745 & 0.020 & 0.019 & $-$ \\
        
    \bottomrule
    \end{tabular}
    \caption{Results for setting where $X$ has an effect on the NBS. Columns represent the probability of being censored, sample size, WTP ($\lambda$), value of $X$, true NBS ($\theta(\lambda|X)$), mean point estimate (Est.), mean estimated standard error ($\widehat{\text{SE}}$), empirical standard error (ESE), and proportion of simulations in which the null hypothesis of no effect of $X$ is rejected (Pr. Reject).}
    \label{tab:sim-eff}
\end{table}

\clearpage

\begin{table}
    \centering
    \begin{tabular}{ccccccccc}
    \toprule
        Cens. & Sample Size & $\lambda$ & X &  $\theta(\lambda|X)$ & Est. & $\widehat{\text{SE}}$ & ESE & Pr. Reject  \\
    \midrule
    \midrule
        10\% & 500 & 2 & 0 & 0.353 & 0.355  & 0.028 & 0.029 & 0.054 \\
        10\% & 500 & 2 & 1 & 0.353 & 0.355 & 0.047 & 0.049 & $-$ \\
        10\% & 500 & 12 & 0 & 0.527 & 0.528  & 0.029 & 0.029 & 0.048 \\
        10\% & 500 & 12 & 1 & 0.527 & 0.530 & 0.047  & 0.049 & $-$ \\
        10\% & 5000 & 2 & 0 & 0.353 & 0.354  & 0.011 & 0.010 & 0.048   \\
        10\% & 5000 & 2 & 1 & 0.353 & 0.353 & 0.018 & 0.017 & $-$ \\
        10\% & 5000 & 12 & 0 & 0.527  & 0.528 & 0.011 & 0.010 & 0.046    \\
        10\% & 5000 & 12 & 1 & 0.527 & 0.527 & 0.018 & 0.018 & $-$ \\
        30\% & 500 & 2 & 0 & 0.353 & 0.353 & 0.033 & 0.034 & 0.040 \\
        30\% & 500 & 2 & 1 & 0.353  & 0.356 & 0.055  & 0.054 & $-$ \\
        30\% & 500 & 12 & 0 & 0.527 & 0.526 & 0.032 & 0.035 & 0.046 \\
        30\% & 500 & 12 & 1 & 0.527 & 0.529 & 0.053 & 0.051 & $-$ \\
        30\% & 5000 & 2 & 0 & 0.353 & 0.352 & 0.012  & 0.013 & 0.054  \\
        30\% & 5000 & 2 & 1 & 0.353 & 0.353 & 0.021 & 0.020 & $-$ \\
        30\% & 5000 & 12 & 0 & 0.527 & 0.525 & 0.012 & 0.012 & 0.054 \\
        30\% & 5000 & 12 & 1 & 0.527 & 0.527 & 0.020 & 0.020 & $-$ \\
        50\% & 500 & 2 & 0 & 0.353 & 0.354 & 0.043 & 0.041 & 0.046 \\
        50\% & 500 & 2 & 1 & 0.353  & 0.356 & 0.069 & 0.068 & $-$ \\
        50\% & 500 & 12 & 0 & 0.527 & 0.528 & 0.038 & 0.037 & 0.040 \\
        50\% & 500 & 12 & 1 & 0.527 & 0.529 & 0.062 & 0.061 & $-$ \\
        50\% & 5000 & 2 & 0 & 0.353 & 0.354 & 0.017 & 0.016 & 0.036 \\
        50\% & 5000 & 2 & 1 & 0.353 & 0.354 & 0.027 & 0.027 & $-$ \\
        50\% & 5000 & 12 & 0 & 0.527 & 0.527 & 0.014 & 0.014 & 0.054 \\
        50\% & 5000 & 12 & 1 & 0.527 & 0.527 & 0.023 & 0.023 & $-$ \\
    \bottomrule
    \end{tabular}
    \caption{Results for setting where $X$ has no effect on the NBS. Columns represent the probability of being censored, sample size, WTP ($\lambda$), value of $X$, true NBS ($\theta(\lambda|X)$), mean point estimate (Est.), mean estimated standard error ($\widehat{\text{SE}}$), empirical standard error (ESE), and proportion of simulations in which the null hypothesis of no effect of $X$ is rejected (Pr. Reject).}
    \label{tab:sim-noeff}
\end{table}

\clearpage

\begin{table}
    \centering
    \begin{tabular}{ccccccccccccc}
    \toprule
     & & &  & \multicolumn{4}{c}{\underline{Survival Confounded}} & & \multicolumn{4}{c}{\underline{Cost Confounded}} \\
        Conf. & $\lambda$ & X & & $\theta(\lambda|X)$ & Est. & $\widehat{\text{SE}}$ & ESE & & $\theta(\lambda|X)$ & Est. & $\widehat{\text{SE}}$ & ESE \\
    \midrule
    \midrule
        low  & 2 & 0 & \hspace{0.1cm} & 0.354 & 0.346 & 0.012 & 0.012 &\hspace{0.1cm} & 0.358 & 0.329 & 0.011 & 0.012 \\
        low  & 2 & 1 & \hspace{0.1cm} & 0.588 & 0.578 & 0.024 & 0.023  &\hspace{0.1cm} & 0.586 & 0.559 & 0.024 & 0.024  \\
        low & 12 & 0 & \hspace{0.1cm} & 0.527 & 0.516 & 0.011 & 0.012 &\hspace{0.1cm} & 0.527 & 0.519 & 0.012 &  0.012 \\
        low  & 12 & 1 & \hspace{0.1cm} & 0.746 & 0.735 & 0.018 & 0.017 &\hspace{0.1cm} & 0.745 & 0.740 & 0.017 &  0.017 \\
        med.  & 2 & 0 & \hspace{0.1cm} & 0.356 & 0.323 & 0.012 & 0.011 &\hspace{0.1cm} & 0.384 & 0.285 & 0.011 & 0.011 \\
        med.  & 2 & 1 & \hspace{0.1cm} & 0.585 & 0.546 & 0.024 & 0.023 &\hspace{0.1cm} & 0.571 & 0.475 & 0.026 &  0.026 \\
        med.  & 12 & 0 & \hspace{0.1cm} & 0.526 & 0.482 & 0.012 & 0.012 &\hspace{0.1cm} & 0.522 & 0.490 & 0.012 &  0.012 \\
        med.  & 12 & 1 &\hspace{0.1cm} & 0.739 & 0.700 & 0.019 & 0.020 &\hspace{0.1cm} & 0.741 & 0.718 & 0.018 &  0.018 \\
        high  & 2 & 0 &\hspace{0.1cm} & 0.362 & 0.287  & 0.011 & 0.011 &\hspace{0.1cm} & 0.423 & 0.251 & 0.011 &  0.011 \\
        high  & 2 & 1 &\hspace{0.1cm} & 0.576 & 0.488 & 0.024 & 0.024 &\hspace{0.1cm} & 0.556 & 0.386 & 0.027 &  0.027 \\
        high  & 12 & 0 &\hspace{0.1cm} & 0.523 & 0.425 & 0.012 & 0.012  &\hspace{0.1cm} & 0.512 & 0.427 & 0.012 &  0.012 \\
        high  & 12 & 1 &\hspace{0.1cm} & 0.722 & 0.631 & 0.020 & 0.020 &\hspace{0.1cm} & 0.720 & 0.651 & 0.020 &  0.021 \\
    \bottomrule
    \end{tabular}
    \caption{Results concerning sensitivity to unmeasured confounding of the exposure-survival and the exposure-cost relationship. Columns represent the WTP ($\lambda$), value of $X$, true NBS ($\theta(\lambda|X)$), mean point estimate (Est.), mean estimated standard error ($\widehat{\text{SE}}$), and empirical standard error (ESE).}
    \label{tab:cnbs-sensitivity}
\end{table}

\begin{table}
    \centering
    \begin{tabular}{cccccc}
    \toprule
          \textbf{Treatment} & $\pmb{\lambda}$ & \textbf{Charlson(0)} & \textbf{Charlson(1)} & \textbf{Charlson(2+)}  \\
    \midrule
    \midrule
       \underline{\textbf{Stage I}}& & & &  \\
        RT & 50  & 0.54 (0.51, 0.55) &  0.52 (0.50, 0.54) &  0.51 (0.48, 0.53)\\
           & 120  & 0.55 (0.52, 0.62) & 0.53 (0.52, 0.56) &  0.53 (0.50, 0.55) \\
        CT & 50  & 0.63 (0.56, 0.68) & 0.60 (0.53, 0.66) &  0.56 (0.49, 0.63) \\
            & 120 & 0.67 (0.60, 0.71) & 0.64 (0.58, 0.70) & 0.60 (0.54, 0.67) \\
            & & & & \\
        \underline{\textbf{Stage II}}    & & & &  \\
        RT & 50  & 0.59 (0.54, 0.64) & 0.57 (0.54, 0.63) & 0.56 (0.52, 0.62) \\
           & 120  & 0.62 (0.56, 0.65) & 0.60 (0.56, 0.65) & 0.60 (0.54, 0.64) \\
        CT & 50  & 0.58 (0.37, 0.69) & 0.54 (0.34, 0.67) &  0.49 (0.30, 0.63)\\
            & 120 & 0.64 (0.45, 0.74) & 0.61 (0.43, 0.72) & 0.58 (0.40, 0.70)) \\
    \bottomrule
    \end{tabular}
    \caption{Estimated NBS comparing RT to control and CT to control conditional on Charlson index and cancer stage. Non-parametric bootstrap-based $95\%$ CIs for each estimated NBS are provided in parentheses.}
    \label{tab:nbs_charlson_stage}
\end{table}

\clearpage

\clearpage

\end{document}